\documentclass[conference]{IEEEtran}
\IEEEoverridecommandlockouts
\usepackage{cite}
\usepackage{amsmath,amssymb,amsfonts}
\usepackage{algorithmic}
\usepackage{graphicx}
\usepackage{textcomp}
\usepackage{xcolor}
\usepackage[multiple]{footmisc}
\usepackage{caption}
\usepackage{subcaption}
\usepackage{color}
\usepackage{eso-pic}
\usepackage{numprint}

\usepackage[pdftex]{hyperref}
\usepackage{tikz}

\newcommand\copyrighttext{%
    \footnotesize \textcopyright 2021 IEEE. Personal use of this material is permitted.
    Permission from IEEE must be obtained for all other uses, in any current or future
    media, including reprinting/republishing this material for advertising or promotional
    purposes, creating new collective works, for resale or redistribution to servers or
    lists, or reuse of any copyrighted component of this work in other works.
    DOI: \href{https://doi.org/10.1109/BigData52589.2021.9671519}{https://doi.org/10.1109/BigData52589.2021.9671519}}
\newcommand\copyrightnotice{%
    \begin{tikzpicture}[remember picture,overlay]
        \node[anchor=south,yshift=10pt] at (current page.south) {\fbox{\parbox{\dimexpr\textwidth-\fboxsep-\fboxrule\relax}{\copyrighttext}}};
    \end{tikzpicture}%
}

\def\BibTeX{{\rm B\kern-.05em{\sc i\kern-.025em b}\kern-.08em
    T\kern-.1667em\lower.7ex\hbox{E}\kern-.125emX}}

\begin{document}

\title{Tarema: Adaptive Resource Allocation for Scalable Scientific Workflows in Heterogeneous Clusters\\}

\author{
    \IEEEauthorblockN{Jonathan Bader\IEEEauthorrefmark{1}, Lauritz Thamsen\IEEEauthorrefmark{1}, Svetlana Kulagina\IEEEauthorrefmark{3},\\
        Jonathan Will\IEEEauthorrefmark{1}, Henning Meyerhenke\IEEEauthorrefmark{3}, and Odej Kao\IEEEauthorrefmark{1}}

    \IEEEauthorblockA{
        \IEEEauthorrefmark{1}
        \{jonathan.bader, lauritz.thamsen, will, odej.kao\}@tu-berlin.de, Technische Universität Berlin, Germany\\
    }
    \IEEEauthorblockA{
        \IEEEauthorrefmark{3}
        \{kulagins, meyerhenke\}@hu-berlin.de, Humboldt-Universität zu Berlin, Germany\\
    }

}
\maketitle
\copyrightnotice
\pagestyle{plain}

\begin{abstract}
    Scientific workflow management systems like Nextflow support large-scale data analysis by abstracting away the details of scientific workflows.
    In these systems, workflows consist of several abstract tasks, of which instances are run in parallel and transform input partitions into output partitions.
    Resource managers like Kubernetes execute such workflow tasks on cluster infrastructures.
    However, these resource managers only consider the number of CPUs and the amount of available memory when assigning tasks to resources; they do not consider hardware differences beyond these numbers, while computational speed and memory access rates can differ significantly.

    We propose \emph{Tarema}, a system for allocating task instances to heterogeneous cluster resources during the execution of scalable scientific workflows.
    First, Tarema profiles the available infrastructure with a set of benchmark programs and groups cluster nodes with similar performance.
    Second, Tarema uses online monitoring data of tasks, assigning labels to tasks depending on their resource usage.
    Third, Tarema uses the node groups and task labels to dynamically assign task instances evenly to resources based on resource demand.
    Our evaluation of a prototype implementation for Kubernetes, using five real-world Nextflow workflows from the popular nf-core framework and two 15-node clusters consisting of different virtual machines, shows a mean reduction of isolated job runtimes by 19.8\% compared to popular schedulers in widely-used resource managers and 4.54\% compared to the heuristic SJFN, while providing a better cluster usage.
    Moreover, executing two long-running workflows in parallel and on restricted resources shows that Tarema is able to reduce the runtimes even more while providing a fair cluster usage.
\end{abstract}

\begin{IEEEkeywords}
    Resource Management, Scientific Workflows, Profiling, Heterogeneous Cluster Resources, Scheduling
\end{IEEEkeywords}

\section{Introduction}\label{sec:INTRO}
Scientific workflow management systems (SWMS) like Nextflow~\cite{nextflow}, Pegasus~\cite{pegasus}, or Snakemake~\cite{koster2012snakemake} help scientists to abstract, compose and execute scalable data analysis processes in domains such as bioinformatics, geosciences, or physics~\cite{bux2013parallelization,deelman2019evolution,workflowcharacterization}.
In these systems, workflows consist of several abstract tasks, of which numerous data-parallel instances run simultaneously on cluster nodes and transform input data partitions into output partitions.
With the help of resource managers like Kubernetes~\cite{kubernetes}, Slurm~\cite{slurm}, HTCondor~\cite{condor} or Yarn~\cite{vavilapalli2013apache} tasks are scheduled to the infrastructure components.
This is necessary since workflows can consist of a large number of tasks, which are often recurring~\cite{witt2019feedback,workflowcharacterization,cybershake}.
This specific pattern, combined with large amounts of data, leads to long runtimes on clusters such as several days or weeks~\cite{witt2019feedback,cybershake}.
Therefore, data-parallel computing on large scale-outs is needed to increase the throughput and to ensure that the analysis executes in a certain timeframe.

Scientists often use a heterogeneous cluster infrastructure to run these data analysis workflows.
While some clusters are planned with heterogeneous nodes to support multiple purposes, some get partially upgraded, and in other clusters failed hardware components are replaced with newer ones~\cite{cpuperformance}.
These resources differ in many aspects like age of the components, CPU cores, memory size, or storage speed~\cite{cpuperformance}.
Even the same number of available CPU cores or memory can lead to highly different runtimes, for instance when instructions per cycle or memory clock rates differ.
Considering such heterogeneous characteristics when allocating tasks onto the infrastructure leads to better resource usage and shorter runtimes.
Manually and statically allocating a large number of tasks onto such a heterogeneous infrastructure is impractical.
The above mentioned resource managers perform the allocation and assignment while ensuring that the task resource requirements are met.
However, resource managers treat each task itself as a black box.
In addition, fine-grained heterogeneity aspects, beyond the amount of cores and memory sizes, are not taken into account.
This leads to a suboptimal allocation, where task and infrastructure profiles are not considered.

This paper presents Tarema, a system addressing this problem through infrastructure profiling and the use of workflow monitoring data, which support the adaptive allocation of resources for scientific workflow tasks in heterogeneous cluster infrastructures.
The infrastructure profiling analyzes the available computing infrastructure regarding performance metrics and their available hardware with a set of microbenchmarks.
Then, Tarema determines node groups through clustering nodes with similar performance metrics.
Based on the profiling results, we are able to label the nodes inside our infrastructure according to the performance characteristics.
In the second step, Tarema uses the monitoring data of the hardware usage of all executed tasks at runtime.
This information is then used to create a performance profile and label recurring tasks.
In the third step, Tarema conducts a scoring which uses the labels from profiling and monitoring to allocate task instances to resources.
These data are then used by the scheduler of the resource manager to assign the task instances to the managed nodes.
Tarema combines the named steps in a system, which aims to fill the gap between the SWMS and the resource manager.

\textit{Contributions.} The contributions of this paper are:

\begin{itemize}
    \item The approach Tarema, which uses infrastructure profiling and task monitoring data to create task-resource profiles, which are then used to determine a task-resource allocation through matching the profiles.
    \item A practical implementation of Tarema, with a cluster profiling tool\footnote{Available at github.com/CRC-FONDA/tarema-cluster-profiler} for heterogeneous infrastructures, an extended version of the SWMS Next\-flow\footnote{Available at github.com/CRC-FONDA/tarema-nextflow-extension}, and a custom Kubernetes scheduler\footnote{Available at github.com/CRC-FONDA/tarema-k8-scheduler}.
    \item An evaluation of our prototype on two heterogeneous cluster infrastructures by using five real-world scientific workflows from the popular nf-core framework\footnote{github.com/nf-core/} and a comparison of our system with three frequently used scheduling approaches for resource managers and the heuristic SJFN.
    \item Our evaluation shows that Tarema helps SWMS to decrease workflow runtimes significantly while providing a lower variance and a better resource utilization.
\end{itemize}

\section{Problem Statement}\label{sec:PROBLEM_DESCRIPTION}
This section first explains the execution model of scalable scientific workflows that we assume in this paper.
We further specify our problem statement and the assumptions we make.

\paragraph{Execution model} Formally, a workflow $W(T,E)$ is a directed acyclic graph (DAG), which consists of a set of tasks $T$ and a set of directed edges $E$.
An edge represents a dependency between two tasks, as well as the order of their execution.
The predecessor task, the one on the outgoing side of the edge, always has to finish executing before the successor task can initiate work.
The input data of a single task definition can be divided into several parts to create multiple task instances which can run in parallel.
Tasks transform these data to generate output partitions used by successor tasks.
We assume that the tasks communicate via files.

\begin{figure}
    \includegraphics[width=\columnwidth]{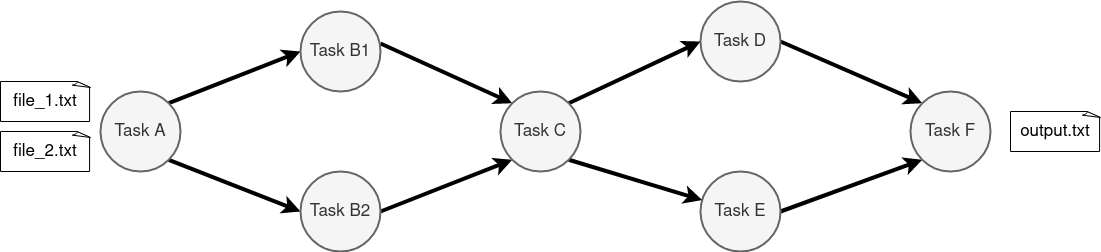}
    \caption{Workflow execution model with 7 tasks instances, two input files and one output file.}
	\label{fig:execution_model}
\end{figure}

As an example, Figure~\ref{fig:execution_model} shows a workflow with 7 task instances and two input files.
After task A finished, the workflow performs a fork and executes two instances of task B in parallel.
Task C joins the results, while the succeeding tasks D and E run in parallel.
The task without an outgoing edge, task F, merges the results from task D and E and produces the output file.

SWMS like Nextflow~\cite{nextflow}, Pegasus~\cite{pegasus}, or Snakemake~\cite{koster2012snakemake} take workflow descriptions, parse them, send the task instances one-bye-one to the resource manager and supervise their execution, making sure that their dependencies are not violated~\cite{bux2013parallelization, saasfee}.
Resource managers like Kubernetes~\cite{kubernetes}, Slurm~\cite{slurm}, HTCondor~\cite{condor} or Yarn~\cite{vavilapalli2013apache} receive task instances and treat them as black-boxes~\cite{witt2019predictive, witt2019feedback}, have no knowledge about task graphs, actual hardware usage, or runtimes  and can therefore only apply simple scheduling approaches.
Simple scheduling means assigning task instances to resources based on amount of resources reserved by users and available on nodes, such as cores and memory, without information on the tasks' actual resource demands and node performance.
For instance, Kubernetes uses a scheduler that works in a round-robin fashion\footnote{kubernetes.io/docs/concepts/scheduling-eviction/scheduler-perf-tuning}, while several YARN distributions use a fair scheduling approach\footnote{docs.datafabric.hpe.com/62/AdministratorGuide/Job-Scheduling.html}\footnote{bdlabs.edureka.co/static/help/topics/admin\_fair\_scheduler.html}.
We therefore define the following problem statement:

\paragraph{Problem statement}

\emph{How to dynamically allocate specific resources in heterogeneous commodity cluster infrastructures for black-box task instances of scalable scientific workflows based on actual hardware performance and resource demand?}

\paragraph{Assumptions} Addressing the problem statement, we make the following assumptions:

\begin{enumerate}
	\item[A1:] A heterogeneous cluster environment where nodes not only differ in the amount of cores, memory, or disks, but also in performance characteristics of these resources.
	\item[A2:] During the initial profiling phase, the respective cluster nodes do not run any workloads to avoid interferences.
	\item[A3:] Workflows are executed repeatedly with possibly different input data.
	\item[A4:] The scientific workflow management system uses the previously described execution model.
\end{enumerate}

\section{Related Work}\label{sec:RELATED_WORK}
Several categories of works are related to our approach.

\paragraph{Scheduling Scientific Workflows on Heterogeneous Clusters}

Scheduling workflows can be done in a static or dynamic manner~\cite{taxonomyscheduling}.
Static scheduling addresses the problem of assigning a set of tasks to compute resources in advance.
The Heterogeneous Earliest Finish Time (HEFT)~\cite{heft} is a frequently cited static heuristic scheduling approach.
The algorithm, however, requires comprehensive knowledge about execution times between all task resource pairs, communication times between dependent tasks, and the DAG.
Dynamic scheduling approaches are in turn more flexible and map tasks to resources at runtime.
There exist many dynamic scheduling methods, where some are based on static scheduling approaches and consider heterogeneous clusters, like P-HEFT~\cite{pheft}.
Due to the extensive knowledge which is required to use many of these approaches, they are often not feasible in real-world systems.
Our proposed system uses profiling and workflow monitoring data to estimate infrastructure characteristics and task-resource demands, which are then used to dynamically allocate resource to tasks.
The information about the task-resource allocation can then be used by schedulers that determine the order of task instance executions and exact placement on clusters

In addition, workflow scheduling distinguishes between scheduling of a single isolated workflow and multiple workflows~\cite{schedulingworkflows}.
Many scheduling approaches are not feasible for scheduling multiple workflows.
Especially if the exact number of executed workflows and the order of tasks are unknown, only dynamic methods are an option.
Tarema can be configured to support the allocation of isolated and multiple workflows.
Since we assume that clusters are shared and the single isolated workflows are not the only workload, even the scheduling of isolated workflows aims to achieve an even distribution of tasks to the infrastructure according to the capabilities.

\paragraph{Scheduling of scalable workflow and batch jobs}Paragon~\cite{paragon} or Quasar~\cite{quasar} propose the use of profiling to determine resource allocation for workloads.
These systems also assume that there is no knowledge about workload runtimes and infrastructure characteristics.
Therefore, Paragon uses short profiling runs to classify workloads and continuously recommends a target according to the characteristics.
Quasar extends this model by integrating scale-out and scale-up impact estimations.
Paragon and Quasar are full resource management systems, addressing both resource allocation and assignment, using machine learning methods to schedule batch and continuous user-facing tasks in heterogeneous clusters.
In contrast, Tarema's scope is resource allocation, so it can be used with existing resource management systems and different schedulers.

Rupam~\cite{xu2018heterogeneity} considers the resource usage of tasks and hardware characteristics to schedule Spark tasks onto heterogeneous resources.
Similar to Tarema, the authors consider not only the number of CPU cores, the amount of memory, or disk size but also different speeds.
However, Rupam's node monitoring metrics are partly static and not sufficient to fully model heterogeneity.
In Tarema, we resolve this problem through extending static attributes with microbenchmarks.
Tarema groups resources, which leads to a fair resource usage and is not conducted by Rupam.
Additionally, Rupam is designed for Spark's execution model, while Tarema can run with different SWMS.

StarPU~\cite{starPU} is a platform that uses a dynamic scheduling approach to schedule tasks to CPU and GPU nodes and optimizes data movements along scheduling tasks.
StarPU does not require prior knowledge about task makespan and offers an API to design custom schedulers on a low abstraction level.
In StarPU user have to provide a cost model for each task.
While StarPU is used in the domain scheduling in high-performance computer architectures, Tarema's scope is resource allocation for scientific workflow tasks in commodity clusters.
Additionally, Tarema does not rely on a user defined cost model for tasks.
Instead, Tarema uses infrastructure profiling, the monitoring data from the SWMS, and relative performance scores as well as group labels to influence allocations.

\paragraph{Cluster Configuration for Distributed Batch Processing} Profiling-based approaches use probe jobs on sample data to determine a near-optimal resource configuration for a given workload~\cite{cherrypick,micky,arrow}.
These samples run on different machines with the goal of estimating a near-optimal cluster configuration.
The key techniques of these existing profiling-based approaches for cluster configuration can be transferred and applied to facilitate adaptive task-resource matching.
Profiling tasks in the domain of scientific workflows has drawbacks.
Some tasks have short runtimes of even less than a few seconds, while others run for hours and days ~\cite{workflowcharacterization}.
Our approach abandons task profiling and instead conducts an infrastructure profiling once, then uses the task performance data from the monitoring at runtime.

Some systems make use of historic runtime data to model the scale-out behavior of a given job~\cite{bell, will2020towards, ernest, will2021c3o}.
Through the provisioning of shared runtime data, performance profiles of identical jobs can be used to create models prior to execution.
Since scientific workflows typically run for multiple days and with different input datasets, it is usually too costly to await the availability of historic runtime data before optimizing.
In comparison, Tarema creates task performance profiles already at workflow runtime.
Also, Tarema neither aims to estimate job runtimes nor to configure entire cluster reservations.

\paragraph{Task Runtime Prediction} Other approaches aim to predict the runtime of tasks~\cite{witt2019predictive, wu2011performance, nadeem2009using}.
Based on historic data, different prediction models can be applied.
Tarema does not intent to predict the runtime.
Instead, our aim is to estimate task-resource demands in relation to other workflow tasks to create task performance profiles.
Thereby, we are not dependent on a certain number of workflow runs to train the models, nor on the input data size since the overall task-resource profile stays constant.

\paragraph{Grouping of Resources and Tasks}Infrastructure clustering is frequently used~\cite{triplet,lee2011heterogeneity,lin2011scheduling} and aims to build homogeneous groups of machines, to take advantage of this at resource allocation time.
Triplet\cite{triplet} is a scheduling algorithm that clusters groups of machines with similar characteristics.
In addition to machine grouping, Triplet performs clustering of tasks into groups with the goal to reduce communication times.
Since the resource manager has no knowledge about the DAG, Tarema conducts no task clustering.
Tarema uses infrastructure grouping to identify similar machines, which are then used to allocate the tasks according to the groups' capabilities.

\section{Approach}\label{sec:APPROACH}
This section first presents an overview about our system.
It then explains the three phases: \textcircled{\raisebox{-0.9pt}{1}} cluster profiling, \textcircled{\raisebox{-0.9pt}{2}} dynamic task monitoring and labeling, and \textcircled{\raisebox{-0.9pt}{3}} task to resource allocation.

\subsection{Overview}
\label{subsec:overview}

\begin{figure*}
    \includegraphics[width=\textwidth]{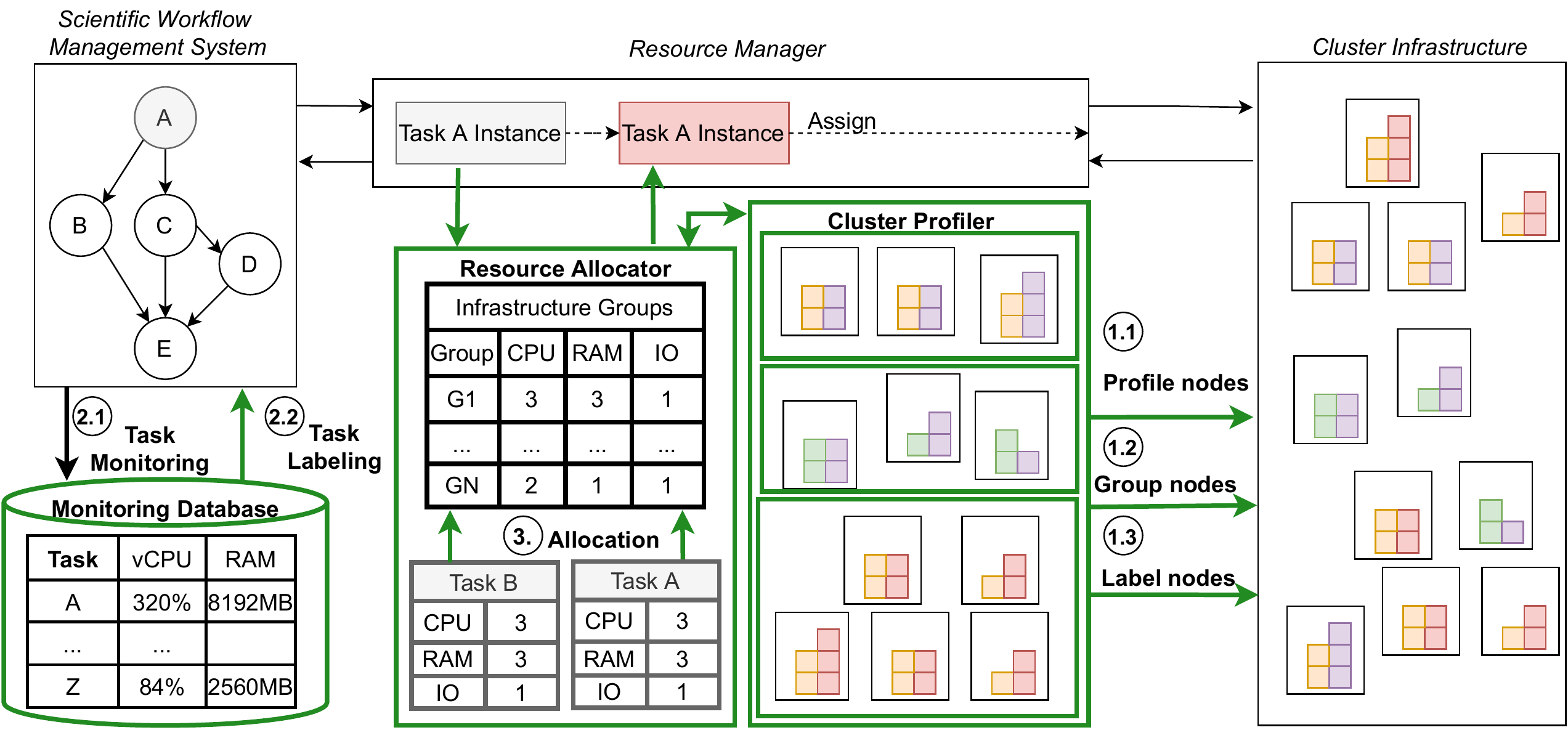}
    \caption{Overview of a typical scientific workflow environment, extended by our Tarema approach (green).} \label{fig:arch}
\end{figure*}

Tarema is a system which aims to bridge the gap between the SWMS and the resource manager through allocating tasks to heterogeneous resources based on extensive infrastructure knowledge and automatically created task performance labels.
Figure~\ref{fig:arch} gives an overview about our approach, where the bold boxes and lines are part of Tarema.
Phase \textcircled{\raisebox{-0.9pt}{1}} conducts the infrastructure profiling.
In Step \textit{1.1}, the cluster profiler uses a set of microbenchmarks and hardware analysis tools to gather performance and static node characteristics.
In Step \textit{1.2}, based on the profiling information, our approach builds node groups with similar performance profiles.
Then, in Step \textit{1.3}, Tarema assigns node labels that express performance and static characteristics and can be used by the resource managers during the allocation phase.
In \textcircled{\raisebox{-0.9pt}{2}}, Tarema uses the online monitoring data from the SWMS and labels tasks at runtime through a workflow management system extension.
The monitoring data in Step \textit{2.1} contain task-resource demands and statistics about active and historic workflows.
During the task-labeling Step \textit{2.2}, these data are used to build a task performance profile for recurring tasks based on the resource demand relative to other tasks.
In the last phase \textcircled{\raisebox{-0.9pt}{3}}, Tarema conducts dynamic task-resource allocation through the information from the previous steps.
First, an allocation function determines the score between the infrastructure node groups and the task-resource demands through using the labels previously assigned.
Our allocator uses a minimization function to determine a priority list of task-resource allocations.
Then, a scheduling algorithm can use this information to perform the task-resource assignment.

\subsection{Cluster Profiler}
\label{subsec:cluster-profiler}

Since we expect the compute cluster to consist of nodes with different kinds of hardware, we want to gather detailed performance characteristics.
Tarema analyses static node properties like the number of CPU cores, CPU cache sizes, or RAM speed, as well as dynamic performance metrics.
To this end, our approach uses microbenchmarks to measure CPU, memory and I/O characteristics.
This step can be executed in parallel, takes less than a minute, and is only done once for each node.
To consider hardware changes and failures in the cluster, one can run the profiler once changes in the cluster are identified by the resource manager.
By using a provisioning and configuration management tool, the cluster profiler ensures that the required dependencies are installed on all target machines.

Before labeling the nodes depending on the gathered performance metrics, Tarema builds node groups with similar performance characteristics.
For this, we use a clustering algorithm that uses a control function, like the silhouette score~\cite{kaufman2009finding}, to estimate the number of node groups.
The features we select for the clustering will later result in labels.
In the default settings we use features for the CPU speed, the memory speed, sequential read/write performance and random read/write performance.
The features can be individually selected and extended to fulfill specific task-resource mapping goals.
A possible extension could be a certain CPU flag, the integration of hardware accelerators or labels for GPU instances.
After building the node similarity groups, Tarema estimates the order of groups for all features, where weaker performance results in a lower rank.
Then, depending on the node's rank, we map the respective labels to scalar values ranging from 1 to $n$, where $n$ is the number of node groups.

These labels are assigned to the nodes controlled by the resource manager and can then be used during the allocation process.

\subsection{Dynamic Task Monitoring and Labeling}
\label{subsec:task-profiler}

Normally, the workflow system would directly submit the tasks to the resource managers, to allocate these for execution on the infrastructure.
We intercept this process through extended task monitoring and labeling.

The online monitoring data of the runtime metrics are retrieved from the SWMS monitoring capabilities, which mostly rely on an operating systems monitor.
A database extends the monitoring and stores the executed tasks together with their runtime metrics and related workflows.
In addition, statistics about currently executed and historical workflows are maintained and updated with the execution of new tasks.
Depending on the number of node groups and the amount of respective nodes in the group, Tarema creates task labels at runtime.

Let us illustrate the labeling for the CPU speed.
Let $G$ be the list of node groups, $g_i$ the $i$-th node group in this list and $n$ the number of elements in $G$.
We denote the total number of CPU cores inside node group $g_i$ as $m_i$.
We sort the elements from $G$ in ascending order depending on the CPU performance score.
Then, Tarema creates $n+1$ percentiles $p_i$, where:

\[
    p_0 = 0; \qquad p_i = \frac{m_i}{\sum_{k=1}^{n} m_k} + p_{i-1},  i \in [1,n-1]; \qquad p_n = 1;
\]

The denominator is the sum of all CPU cores in the groups.
The percentiles express the respective feature distribution over the nodes, in this case the number of CPU cores.
The monitoring database contains the CPU utilization monitoring data from tasks of the currently running workflows and their historic executions.
For example, a CPU usage monitoring value for task t of 210\% would indicate a measured full usage of 2 full CPU cores and 10\% of a third one.
We sort the monitoring task data for the respective workflow and feature, in this case the CPU utilization, in ascending order.
Now, Tarema applies the percentiles on the sorted CPU utilization data.
The CPU utilization value $v_{p_n}$ at the bounds of percentiles $p_n$ is used to build $n$ intervals  $ [0, v_{p_1}[,[v_{p_1}, v_{p_2}[,\dots,[ v_{p_{n-1}}, +\infty [$.
An example interval for three node groups could look like the following: $ [0, 54\%[,[54\%, 112\%[,[ 112\%, +\infty [$.
Tarema searches the database for historic data from the task to be submitted.
In the case of existing historic executions, we have an estimation of the resource usage.
Tarema examines in which interval the CPU utilization of the task to be submitted lays.
We then label the CPU feature with a value between 1 to $n$ according to the interval rank in which the task falls, where a higher number expresses higher resource demand.
Tarema conducts this step also for features like RAM (memory speed) or I/O (sequential read/write).
Through creating percentile intervals according to the resource capabilities of node groups, we aim to achieve a fair task distribution so that less demanding tasks do not block the most capable nodes.

\subsection{Adaptive Resource Allocation}
\label{subsec:task-scheduling}

After the task has been submitted to the resource manager, Tarema's resource allocator aims to find the best-fitting resource for each task.
The results from the resource allocation, as well as the provided knowledge about infrastructures and tasks can then be used to implement a scheduling algorithm.
To support the allocation, we create labels for the infrastructure and tasks according to their performance characteristics.
For the resource allocation of our system, we use a scoring algorithm to determine the best match between a task and the available resources.
By calculating a score of the task in relation to each node group, we can create a priority list of node-groups to allocate the task.
Let $S$ be the list of all node-group-task pairs, and $P$ is the list of pairs where nodes inside node-group $n$ satisfy the resource requirements of task $t$.
Our score function for one node-group-task pair is defined as:

\[
    f(n,t) = \sum_{k=1}^{q} |n_k - t_k|,
\]
where $q$ is the number of features (CPU, Memory, I/O), and $n_k$ and $t_k$ are scalar feature labels for the respective node-group-task pair.
The values of the scalar feature labels originate from cluster profiling and task monitoring in the previous steps.
After applying $f(n,t)$ on all pairs of $P$, the minimum list entry determines the near-optimal task-resource allocation.
Table~\ref{tab:task-resource-function} gives an example where task t has to be assigned to one of four available node groups.
The task has the labels $t_1=3$, $t_2=3$, $t_3=2$, where $t_1$ is the CPU label, $t_2$ the memory label, and $t_3$ the I/O label.
The result of applying $f(n,t)$ on all pairs of $P$ is the respective sum of the main diagonal.
In the given example, we would prefer nodes from group four since the sum is the minimum.
In the case where no node insight the preferred group has sufficient resources, the priority list of task-resource pairs can be used by the scheduling algorithm to select the next best-fit.
In our scheduling algorithm, our first-order criterion is to select a node from the preferred node-group list.
If there are several node-groups with the same ranking, we select the most powerful node group, which we determine by summing up all scalar feature values.
Inside the group, we select the node with the currently smallest load as the second-order criterion.
Unknown, and therefore unlabeled tasks, are assigned to the nodes with the least load to achieve a fair distribution.

\begin{table}[]
    \caption{Resource Allocation matrix for task t on four node groups}
    \begin{tabular}{ccccccccccl}
        \multicolumn{1}{l}{}     & \multicolumn{1}{l}{}   & \multicolumn{2}{c}{Node Group 1}                                         & \multicolumn{1}{l}{}            & \multicolumn{1}{l}{}  & \multicolumn{1}{l}{}   & \multicolumn{2}{c}{Node Group 3}                                        & \multicolumn{1}{l}{}            &  \\
        &                        & CPU                             & RAM                             & I/O                             &                       &                        & CPU                             & RAM                             & I/O                             &  \\ \cline{2-5} \cline{7-10}
        \multicolumn{1}{c|}{}    & \multicolumn{1}{c|}{}  & \multicolumn{1}{c|}{1}          & \multicolumn{1}{c|}{1}          & \multicolumn{1}{c|}{1}          & \multicolumn{1}{c|}{} & \multicolumn{1}{c|}{}  & \multicolumn{1}{c|}{1}          & \multicolumn{1}{c|}{1}          & \multicolumn{1}{c|}{2}          &  \\ \cline{2-5} \cline{7-10}
        \multicolumn{1}{c|}{CPU} & \multicolumn{1}{c|}{3} & \multicolumn{1}{c|}{\textbf{2}} & \multicolumn{1}{c|}{-}          & \multicolumn{1}{c|}{-}          & \multicolumn{1}{c|}{} & \multicolumn{1}{c|}{3} & \multicolumn{1}{c|}{\textbf{2}} & \multicolumn{1}{c|}{-}          & \multicolumn{1}{c|}{-}          &  \\ \cline{2-5} \cline{7-10}
        \multicolumn{1}{c|}{RAM} & \multicolumn{1}{c|}{3} & \multicolumn{1}{c|}{-}          & \multicolumn{1}{c|}{\textbf{2}} & \multicolumn{1}{c|}{-}          & \multicolumn{1}{c|}{} & \multicolumn{1}{c|}{3} & \multicolumn{1}{c|}{-}          & \multicolumn{1}{c|}{\textbf{2}} & \multicolumn{1}{c|}{-}          &  \\ \cline{2-5} \cline{7-10}
        \multicolumn{1}{c|}{I/O} & \multicolumn{1}{c|}{2} & \multicolumn{1}{c|}{-}          & \multicolumn{1}{c|}{-}          & \multicolumn{1}{c|}{\textbf{1}} & \multicolumn{1}{c|}{} & \multicolumn{1}{c|}{2} & \multicolumn{1}{c|}{-}          & \multicolumn{1}{c|}{-}          & \multicolumn{1}{c|}{\textbf{0}} &  \\ \cline{2-5} \cline{7-10}
        &                        & \multicolumn{2}{c}{Node Group 2}                                         &                                 &                       &                        & \multicolumn{2}{c}{Node Group 4}                                        &                                 &  \\ \cline{2-5} \cline{7-10}
        \multicolumn{1}{c|}{}    & \multicolumn{1}{c|}{}  & \multicolumn{1}{c|}{2}          & \multicolumn{1}{c|}{2}          & \multicolumn{1}{c|}{3}          & \multicolumn{1}{c|}{} & \multicolumn{1}{c|}{}  & \multicolumn{1}{c|}{3}          & \multicolumn{1}{c|}{3}          & \multicolumn{1}{c|}{3}          &  \\ \cline{2-5} \cline{7-10}
        \multicolumn{1}{c|}{CPU} & \multicolumn{1}{c|}{3} & \multicolumn{1}{c|}{\textbf{1}} & \multicolumn{1}{c|}{-}          & \multicolumn{1}{c|}{-}          & \multicolumn{1}{c|}{} & \multicolumn{1}{c|}{3} & \multicolumn{1}{c|}{\textbf{0}} & \multicolumn{1}{c|}{-}          & \multicolumn{1}{c|}{-}          &  \\ \cline{2-5} \cline{7-10}
        \multicolumn{1}{c|}{RAM} & \multicolumn{1}{c|}{3} & \multicolumn{1}{c|}{-}          & \multicolumn{1}{c|}{\textbf{1}} & \multicolumn{1}{c|}{-}          & \multicolumn{1}{c|}{} & \multicolumn{1}{c|}{3} & \multicolumn{1}{c|}{-}          & \multicolumn{1}{c|}{\textbf{0}} & \multicolumn{1}{c|}{-}          &  \\ \cline{2-5} \cline{7-10}
        \multicolumn{1}{c|}{I/O} & \multicolumn{1}{c|}{2} & \multicolumn{1}{c|}{-}          & \multicolumn{1}{c|}{-}          & \multicolumn{1}{c|}{\textbf{1}} & \multicolumn{1}{c|}{} & \multicolumn{1}{c|}{2} & \multicolumn{1}{c|}{-}          & \multicolumn{1}{c|}{-}          & \multicolumn{1}{c|}{\textbf{1}} &  \\ \cline{2-5} \cline{7-10}
    \end{tabular}
    \label{tab:task-resource-function}
\end{table}

This score function helps Tarema to avoid mapping resource-intensive tasks to less powerful hardware resources, and vice versa.
Matching values closer to zero indicates that task-performance profiles map to node characteristics.
Since higher resource demands result in higher scalar feature values, they have an increased impact on the matching process.

\section{Evaluation}\label{sec:EVALUATION}
This section presents our prototype implementation, the experiment setup, as well as the experimental results from running five real-world scientific workflows with different scheduling approaches on two different clusters.
The entire source code to conduct the evaluation is available online.\footnote{Available at github.com/CRC-FONDA/tarema-experiments}

\subsection{Prototype Implementation}
\label{subsec:prototype-implementation}
In accordance with the three phases in our model, this section first presents the profiler, followed by the scientific workflow management system extension and the scheduler. 
\paragraph{Cluster Profiler}

Our system uses \textit{sysbench} to measure the single-threading and multi-threading performance of the machines.
\textit{Sysbench} runs a  CPU benchmark that verifies prime numbers.
We limit the runtime to ten seconds, and the maximum number to verify to 20000.
In addition, we run \textit{lscpu} on all destination systems to collect more specific static CPU information, which can be used for more specific workload schedules.
An example is a certain CPU flag like Advanced Vector Extension, which can improve certain calculations.

The memory benchmark uses \textit{sysbench} as well.
In the memory test, we set the memory block size buffer to one megabyte, the total memory size to 100 gigabytes, and the number of threads to one.
Running the memory benchmark with all available threads would be possible.
However, there are two major problems doing this.
First, the values will be hardly comparable, for example, a node with 4 vCPUs would score less even with a faster physical memory than a node with 8 vCPUs.
Second, the tasks allocate a fix amount of memory.
Therefore, a benchmark value which has been obtained with more than the requested memory will be misleading.
We therefore decided to use one thread to provide a coarse comparability.
Our tool also provides static information by running \mbox{\textit{dmidecode --type 17}} on the target machines.

We test the I/O performance of the target systems with the tool \textit{fio}.
We conduct one benchmark for sequential read-write and one to measure random read-write characteristics.
In both tests, we set the filesize to 4 gigabytes, and use the libaio library for asynchronous access with a depth of 64.
To omit the page cache and the use of the memory, we set the direct parameter to one.
The \textit{lshw -class storage} command provides the static information about the available devices.

We are aware that modern hardware is tailored to achieve high scores in frequently used benchmarks like sysbench and fio.
Therefore, the results may not fully reflect the actual performance of the hardware.
However, we do not aim to exactly estimate the single infrastructure performances in detail.
Benchmarking the cluster hardware has the goal to relatively compare the nodes inside the cluster.

With the knowledge from the benchmarks, Tarema uses the k-means++ algorithm to form the node groups.
The silhouette score is used to determine the quality of the clusters and the nodes inside the groups, as well as the number of similarity groups.
\paragraph{Nextflow Extension}

In the default configuration, Nextflow gathers task runtime information to create a report file in html.
These monitoring data are retrieved from the Linux process stats (ps) and contain information like CPU usage, resident set size, or read/write data.
We intercept this process and collect the task-resource demand in a PostgreSQL database.
Further, Tarema uses materialized views with indexes in the database to update workflow statistics at task completion.
Before Nextflow submits the task to the resource manager, we start labeling the task.
Therefore, we query historical task executions and hit our materialized views.
Since the resource managers need additional labeling information, Tarema extends the resource managers' executor interface.
\paragraph{Kubernetes Scheduler}

Nextflow supports several resource managers, one of the most popular ones is Kubernetes.
The default Kubernetes scheduler schedules in a round-robin fashion.\footnote{kubernetes.io/docs/concepts/scheduling-eviction/scheduler-perf-tuning}
We therefore use the fabric8io K8 library\footnote{github.com/fabric8io/kubernetes-client} to provide a custom Java Kubernetes scheduler that supports three well-known scheduling approaches for resource managers, a Shortest-Job-Fastest-Node scheduler, as well as Tarema's task-resource allocation logic and the scheduling algorithm.
The scheduling approaches are easily exchangeable and extensible to enable the implementation of custom schedulers. 
Our scheduler observes submitted tasks and active nodes using Informers and Watchers.

\paragraph{Usage with other SWMS}

The cluster profiler only requires the machines in the cluster to run POSIX compliant operating systems and does not depend on any SWMS.
Further, the resource allocator expects labeled Kubernetes pods and is therefore also independent of a specific SWMS.
For using Tarema with other SWMS, the respective monitoring part has to be extended by task labeling.
The SWMS monitoring extensions offer similar metrics like resource usage.
However, the different SWMS do not provide uniform interfaces and, therefore, require individual adaptions.

For example, the popular SWMS Pegasus offers a monitoring REST-API, which could be used to label the tasks.
Snakemake supports panoptes, a monitoring service that can also be used to label the tasks before submitting them to the resource manager.

\begin{table}[bt!]
    \centering
    \caption{The cluster configuration for the 5;5;5 cluster.}
    \begin{tabular}{|l|l|l|l|l|l|}
        \hline
        GCP Name & \# Nodes & vCPUs & Memory & Storage    & Bandwidth \\ \hline
        N1   & 5      & 8    & 32 GB  & HDD & 16 Gbps  \\ \hline
        N2   & 5      & 8    & 32 GB  & HDD & 16 Gbps  \\ \hline
        C2   & 5      & 8    & 32 GB  & HDD & 16 Gbps  \\ \hline
    \end{tabular}
    \label{tab:hardware-specifications-cluster-one}
\end{table}

\begin{table}[bt!]
    \centering
    \caption{The cluster configuration for the 5;4;4;2 cluster.}
    \begin{tabular}{|l|l|l|l|l|l|}
        \hline
        GCP Name & \# Nodes & vCPUs & Memory & Storage    & Bandwidth \\ \hline
        E2   & 5               & 6    & 16 GB  & HDD & 8 Gbps  \\ \hline
        N1   & 4               & 6    & 16 GB  & HDD & 10 Gbps  \\ \hline
        N2   & 4               & 8    & 32 GB  & HDD & 16 Gbps  \\ \hline
        C2   & 2               & 16    & 64 GB  & HDD & 32 Gbps  \\ \hline
    \end{tabular}
    \label{tab:hardware-specifications-cluster-two}
\end{table}

\subsection{Cluster Setup}
\label{subsec:cluster-setup}

\begin{table*}[t]
    \caption{The results of Tarema’s profiling runs. The last column shows the node similarity groups Tarema created.}
    \begin{tabular}{|l|l|l|l|l|l|l|l|l|}
        \hline
        \# of Nodes & CPU events/s & RAM MiB/s     & random write IOPS & random read IOPS & sequential write IOPS & sequential read IOPS & \textbf{Group}  \\ \hline
        \multicolumn{8}{|c|}{\textbf{ 5;5;5 Cluster Configuration}} \\ \hline
        5       & 367-384       & 13800-14300           & 107-108  & 102      & 483      & 481 & \textbf{1}      \\ \hline
        5      & 458-468       & 17500-17700           & 107-108  & 102      & 483      & 481 & \textbf{2}      \\ \hline
        5     & 523-525       & 19800-19900           & 107-108  & 102      & 483      & 481  & \textbf{3}    \\ \hline \hline
        \multicolumn{8}{|c|}{\textbf{5;4;4;2 Cluster Configuration}} \\ \hline
        9       & 368-384       & 13100-14200           & 107-108  & 102      & 483      & 481  & \textbf{1}    \\ \hline
        4       & 469-470       & 17700-17800           & 107-108  & 102      & 483      & 481  & \textbf{2}    \\ \hline
        2      & 522-524       & 19800           & 107-108  & 102      & 483      & 481  & \textbf{3}    \\ \hline
    \end{tabular}

    \label{tab:profiling-clusters}
\end{table*}

Since we focus on heterogeneous cluster environments, we use \textit{Terraform}, an open-source infrastructure as code software\footnote{terraform.io}, to build the evaluation setup in a reproducible manner.

For our evaluation, we choose the Google Compute Platform and select the virtual machines to model heterogeneous infrastructure.

We use two different heterogeneous cluster configurations, where each contains in total 15 machines.
Table~\ref{tab:hardware-specifications-cluster-one} gives an overview about the first cluster used for our experiments.
We choose a uniform distribution of nodes, where each node has the same amount of CPUs and memory.
On the first look the hardware characteristics in Table~\ref{tab:hardware-specifications-cluster-one} look homogeneous.
However, the different names indicate different kind of machines.
Our N1 machines are based on Intel Broadwell platforms with a base clock of 2.0 GHz, N2 are Intel Cascade Lake CPUs with a base clock of 2.8 GHz, and C2 are compute-optimized machines based on Intel Scalable Processors (Cascade Lake) with a higher turbo clock of up to 3.8 Ghz.\footnote{cloud.google.com/compute/docs/machine-types}

The specifications of the second cluster are depicted in Table~\ref{tab:hardware-specifications-cluster-two}.
Here not only the hardware specification is heterogeneous, we also choose a different node distribution.
In addition to N1 nodes, we select E2 machines types, which are cost-optimized and are based on Intel Broadwell with a base clock of 2.2 GHz and model the older hardware nodes existing in a cluster, while the C2 machines can be seen as the newly available infrastructure nodes, which are more powerful but expensive and therefore scarce.

Since the nodes in the cluster reside in the same region and zone, they have a latency in the range of $[1 ms, 2 ms]$ between each other.
We do not consider external load on the cluster during the experiments.

The current Nextflow version requires a single persistent volume claim.
Therefore, it is not viable to take advantage of heterogeneous I/O characteristics and explains why we choose the same storage type for all node instances.

\subsection{Workflow Setup}
\label{subsec:workflow-setup}

\begin{figure}[tb!]
    \includegraphics[width=\columnwidth]{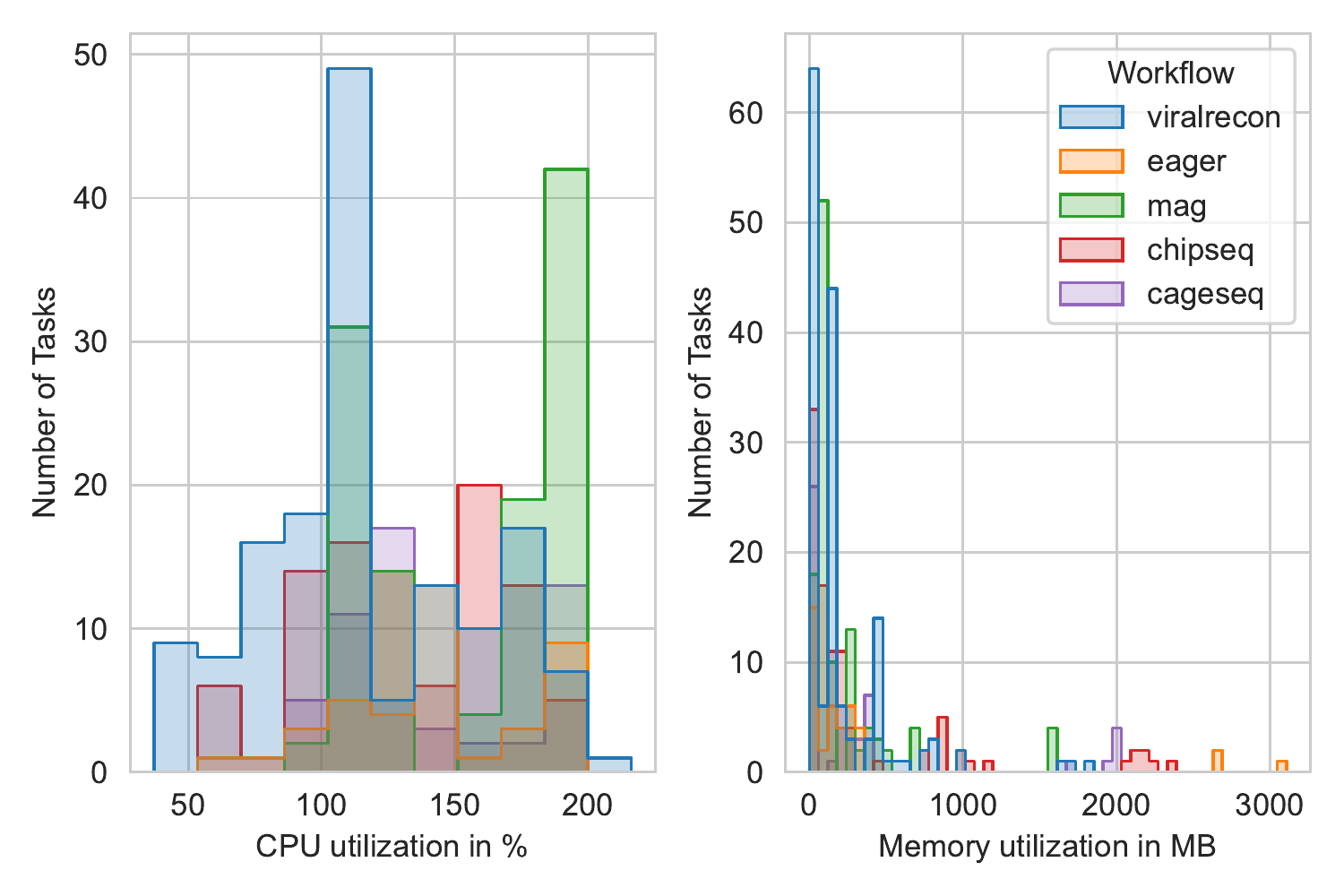}
    \caption{CPU and memory utilization of the experiment workflows.}
    \label{fig:workflow_profiles}
\end{figure}

We chose five publicly available real-world workflows from the popular nf-core repository: Viralrecon - variant calling for viral samples; Eager - ancient DNA analysis; Mag - assembly and binding of metagenomes; CAGE-Seq; Chipseq - peak-calling.
Recall that we mentioned that scalable scientific workflows can run for days and easily exceed terabytes of input data.
We therefore use very small parts of datasets or cut these to reduce the runtimes.
The five chosen workflows have different structures and exhibit diverse resource usage patterns, which can be seen in Figure~\ref{fig:workflow_profiles}.
For example, the Mag workflow contains many CPU-intensive tasks, while Chipseq and Eager contain more memory-intensive tasks.
We assigned all tasks 2 CPUs and 5GB of memory.

\subsection{Cluster Profiling}
\label{subsec:cluster-profiling}

Initially, the Tarema cluster profiler creates a fine-granular infrastructure profile with groups of similar nodes.
Table~\ref{tab:profiling-clusters} shows that Tarema splits the nodes for both configurations into three node groups.
The profiling shows that the hardware performance characteristics inside the respective groups have a small range.
However, there is a significant difference between the node groups' memory and CPU speed.

Tarema then automatically labels the nodes inside the cluster according to the steps described in Section~\ref{sec:APPROACH}.

\begin{figure*}[t]
    \includegraphics[width=\textwidth, keepaspectratio]{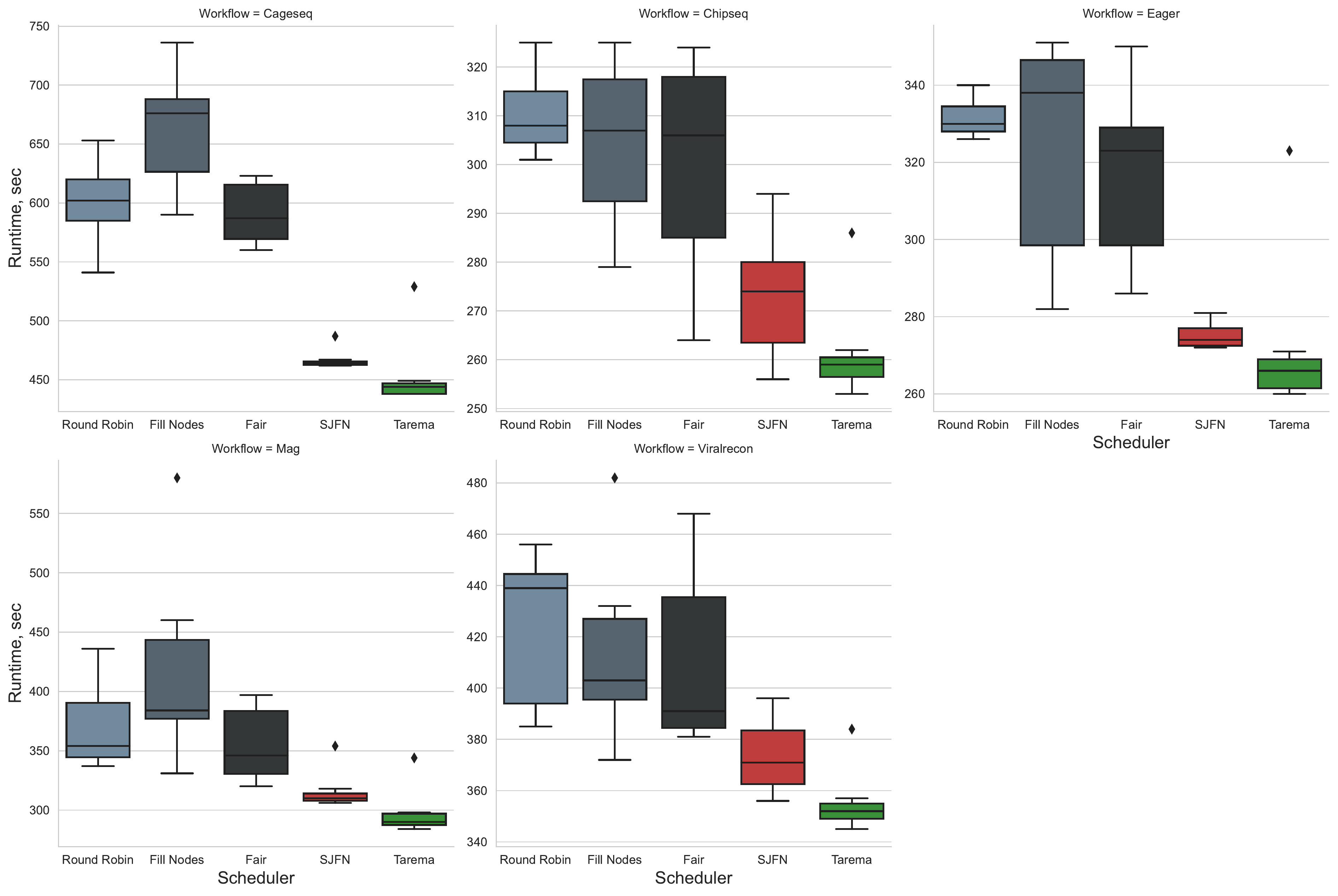}
    \caption{Workflow runtimes on the 5;5;5 cluster comparing Tarema to the baseline schedulers (y-axis not starting at 0, to highlight differences) }
    \label{fig:tarema_555}
\end{figure*}

\begin{figure*}[t]
    \includegraphics[width=\textwidth, keepaspectratio]{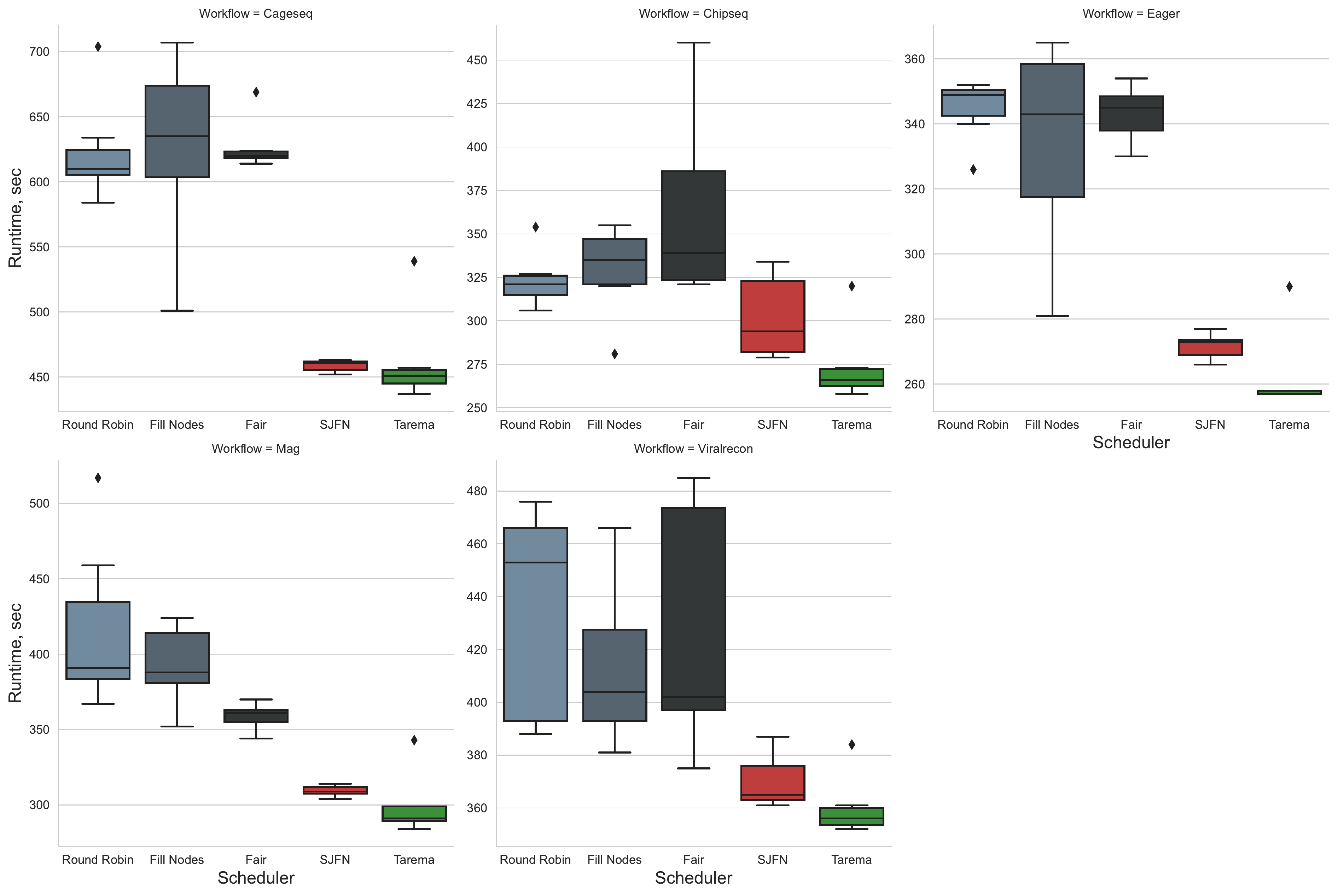}
    \caption{Workflow runtimes on the 5;4;4;2 cluster comparing Tarema to the baseline schedulers (y-axis not starting at 0, to highlight differences)}
    \label{fig:tarema_5442}
\end{figure*}

\subsection{Experiments}

\begin{figure}[h!]
    \includegraphics[width=\columnwidth]{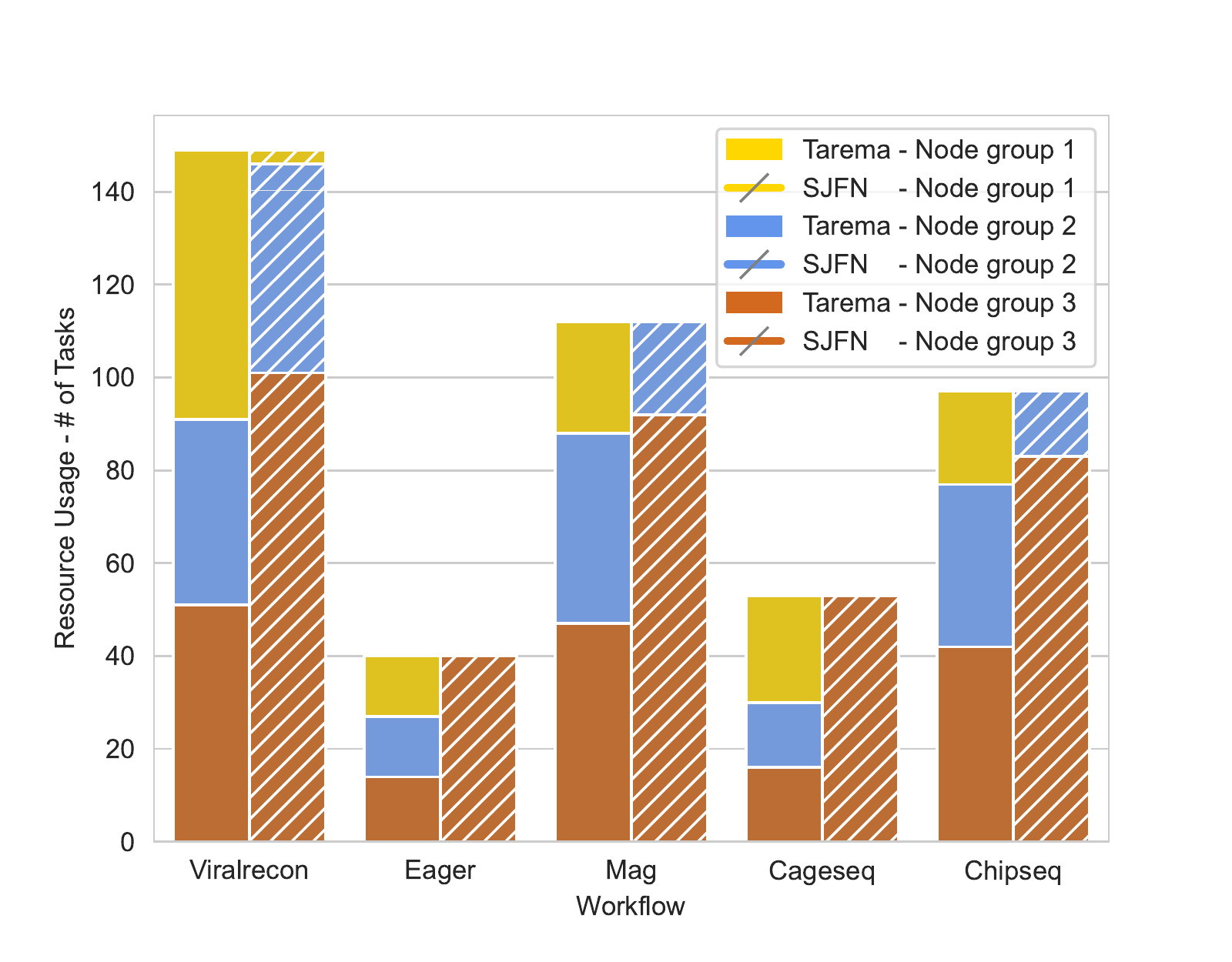}
    \caption{Resource usage Tarema vs. SJFN on cluster 5;5;5.}
    \label{fig:resource_usage_555}
\end{figure}

\begin{figure}[h!]
    \includegraphics[width=\columnwidth]{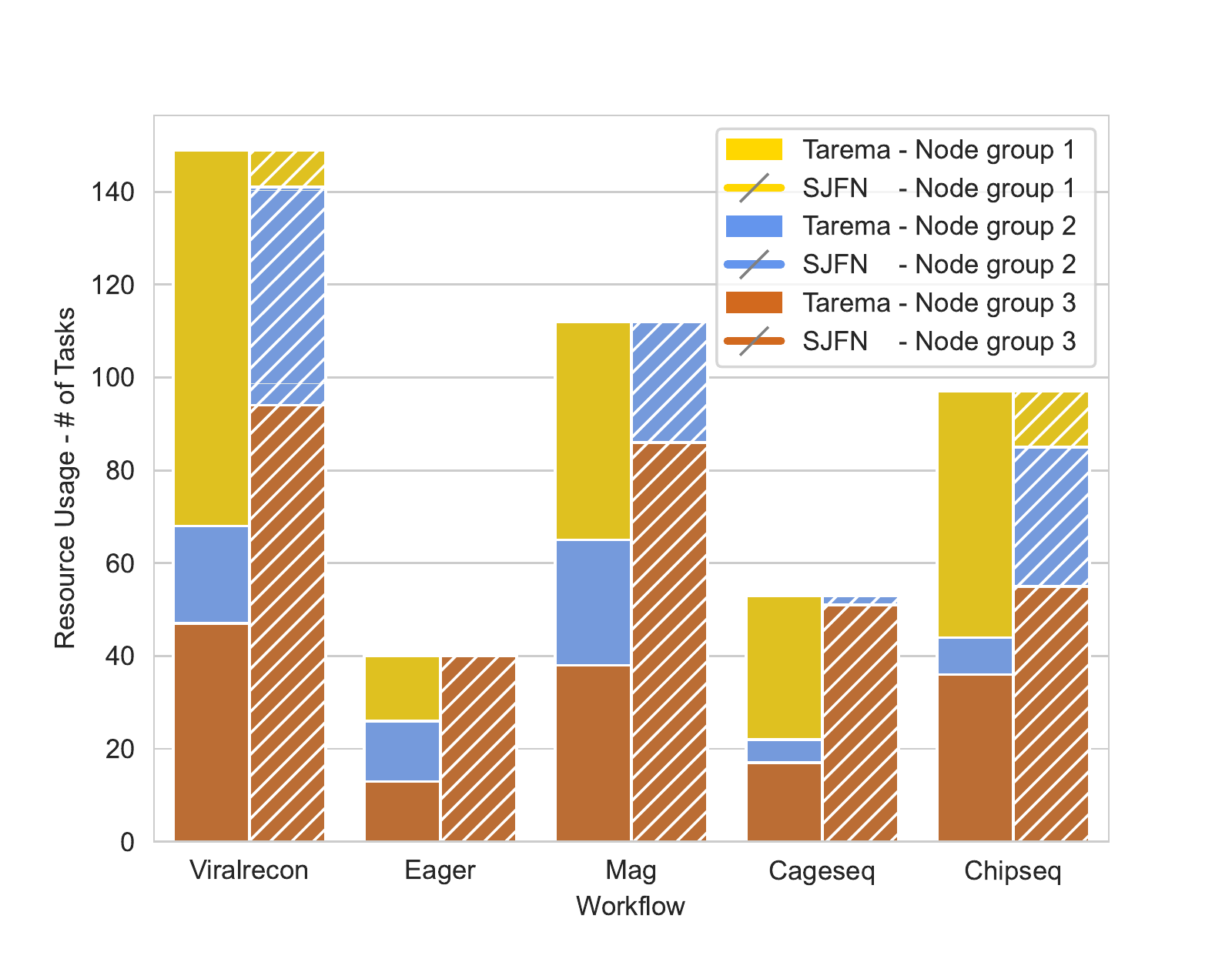}
    \caption{Resource usage Tarema vs. SJFN on cluster 5;4;4;2.}
    \label{fig:resource_usage_5442}
\end{figure}

With our experiments, we aim to show how knowledge about infrastructure performance and task resource demands can help determine task-resource allocations that improve the runtimes of tasks and evenly utilize available resources.
\paragraph{Experiment Design}
First, we examine how three standard scheduling algorithms in widely used resource managers schedule isolated real-world workflows without any infrastructure and task profile knowledge.
As a first set of baselines we selected Round-Robin, Fair Scheduling, and Fill Nodes~\cite{michael2018scheduling}, which are frequently used schedulers in resource managers like Kubernetes, Slurm, or Yarn.
The default Kubernetes scheduler works in a round-robin fashion.
Fair scheduling aims to distribute the resources reserved on all nodes equally and is used similarly in Yarn distributions and Slurm.
The Fill-Nodes scheduling approach aims to fully claim node resources before assigning tasks to the next node in the node list.

Afterwards, we apply the profiling and labeling capabilities of Tarema to gather detailed information about the cluster and the submitted tasks.
With this knowledge, we evaluate Tarema's default allocation and scheduling implementation with regard to workflow runtimes and resource usage.
In addition, we use the obtained knowledge also for the heuristic scheduling approach Shortest-Job-Fastest-Node (SJFN), which is based on Shortest-Job-Next (SJN)~\cite{xhafa2008metaheuristics, schedulingSJFR}.
SJFN is a heuristic scheduling approach which schedules the shortest jobs to the fastest resources with the aim to reduce the turnaround time~\cite{abraham2005rule, schedulingSJFR}.
We include it as another baseline to compare Tarema's performance also against another heuristic scheduling approach that takes into account data on infrastructure and task performance.
Our SJFN implementation uses the historic runtime data from Tarema's monitoring extension to order the tasks for the shortest runtime, as well as the infrastructure data from the profiling phase.
Beyond runtimes, we also compare Tarema and SJFN in the resulting resource utilization.
Later, we compare Tarema with SJFN running multiple workflows.

To ensure that the scheduling results are not influenced by the order of nodes which the scheduler holds in a list, we shuffle the nodes each time.
 
The five schedulers run each workflow seven times on both clusters.
An initial run for each Scheduler-Workflow pair to acquire data dependencies and to pull the docker images is not part of the benchmark.
After the experimental evaluation of each Scheduler-Workflow pair, we delete the database entries.

\paragraph{Experiment results - Isolated Workflows}
The runtimes and their distributions can be seen in the Figures~\ref{fig:tarema_555} and~\ref{fig:tarema_5442}.
One can see that SJFN and Tarema consistently outperform the scheduling approaches which are widely used in resource managers in both clusters and on all workflows.
The experiments in our first cluster result in a geometric mean runtime decrease of 17.87\%, comparing Tarema to the three standard baselines on all workflows.
Meanwhile, in Cluster 5;4;4;2 Tarema is able to decrease the geometric mean runtime by 21.47\%.
In addition, Tarema outperforms the heuristic approach SJFN between 2.1\% and 9.5\% over all setups.
Tarema achieves a 4.65\% runtime decrease in Cluster 5;5;5 and a 4.45\% runtime decrease in Cluster 5;4;4;2 compared to SJFN.

SJFN and Tarema show a lower standard deviation and a lower runtime compared to the standard scheduling approaches.

In general, task runtimes can vary in real-world systems and impact the arrival time of subsequent tasks which might have to be assigned to a less powerful nodes and therefore impact SJFN schedules.
This can affect the runtime through the sequence and timing of tasks.

In contrast to SJFN, the task-resource allocation Tarema conducts is not dependent on the arrival times or the ordering.
However, this only applies as long as there are historic task information that can be used to label the recurring tasks in the workflow.
Figures~\ref{fig:tarema_555} and~\ref{fig:tarema_5442} show that single runtime outliers exist for Tarema executions.
The log data shows that these outliers are the first runs, where little knowledge about tasks is known yet and Tarema has to assign unknown tasks according to fair scheduling.
Therefore, the first runs only partially profit from labeling at workflow runtime.

Figures~\ref{fig:resource_usage_555} and ~\ref{fig:resource_usage_5442} compare the resource usage of Tarema and SJFN for each workflow.
Tarema aims to allocate the tasks according to the existing node group distribution.
However, in contrast to the task-node allocation which Tarema performs based on the resource distribution, SJFN assigns to the most powerful nodes first.

Comparing the resource usage of cluster 5;5;5 with cluster 5;4;4;2 for SJFN, one can see that for the second configuration, more tasks are scheduled to the less powerful node groups two and one.
This behaviour is expected due to fewer C2 and N2 machines in cluster 5;4;4;2, compared to cluster 5;5;5, which are available for allocation.
For Tarema, the resource usage for cluster 5;4;4;2 shows that more tasks are assigned to the least powerful nodes and fewer to the most powerful ones.

In Figures~\ref{fig:tarema_555} and~\ref{fig:tarema_5442} we saw that Tarema outperforms SJFN with regard to runtime.
It might sound counterintuitive that a fair resource allocation achieves lower runtimes compared to a case where only the most powerful nodes are used.
Since SJFN assigns all tasks to the most powerful machines, many tasks have to share the resources.
This sharing can lead to interferences, which can get higher with a higher number of competing tasks~\cite{interferencememory, interferencesgoogle, thamsen2020mary}.

\paragraph{Experiment results - Multiple Workflows}
\label{par:eval_multiple_wfs}

\begin{figure}[h!]
    \includegraphics[width=\columnwidth]{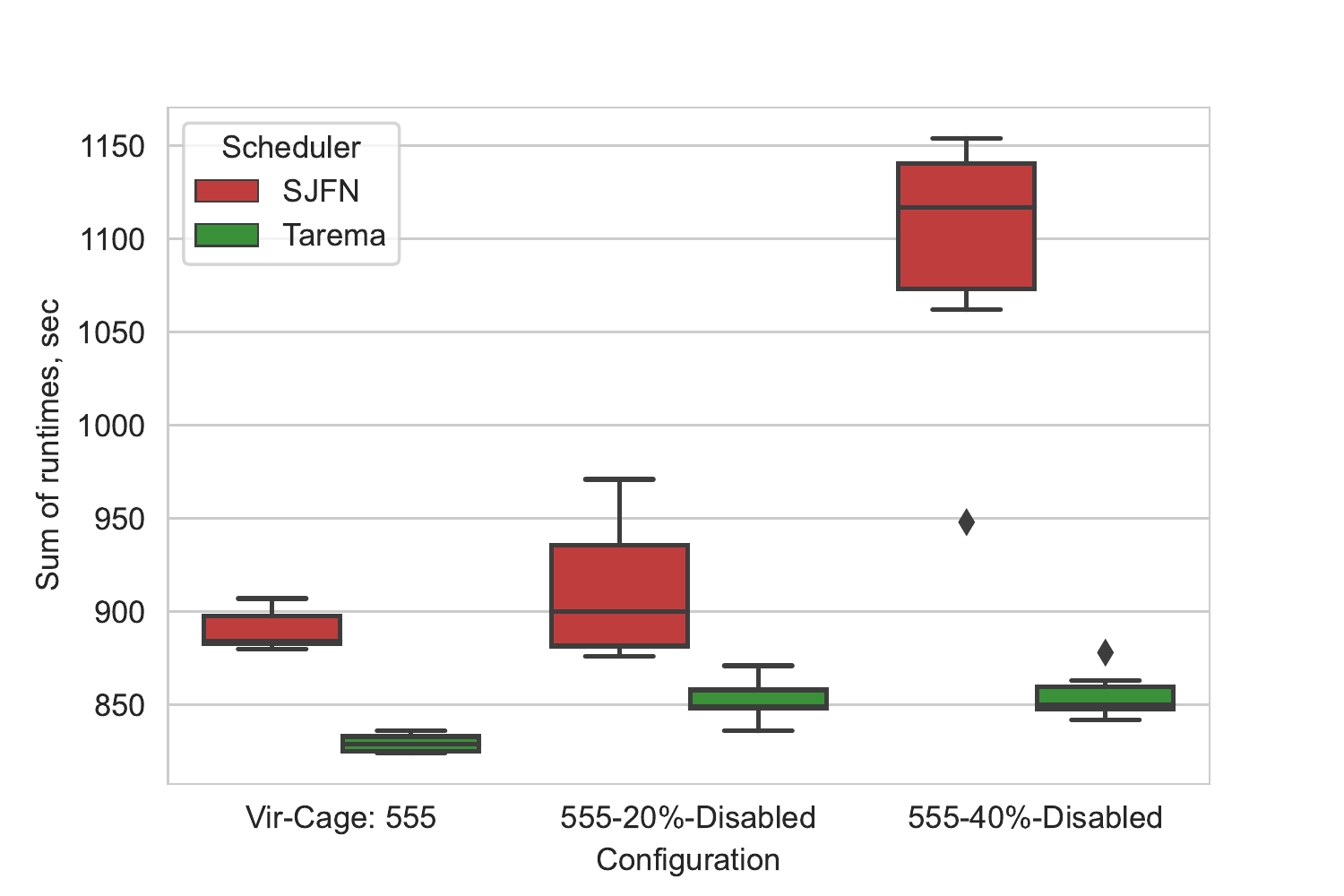}
    \caption{Sum of the workflow runtimes Tarema vs. SJFN.}
    \label{fig:tarema_multiple_wfs}
\end{figure}

The first configuration in Figure~\ref{fig:tarema_multiple_wfs} compares the runtimes of Tarema to SJFN, running Viralrecon and Cageseq, which are the workflows with the longest runtimes, in parallel on the 5;5;5 cluster.
Running Viralrecon and Cageseq in parallel using the first configuration yields a mean runtime decrease of 6.22\%.
In the restricted configuration, we disabled 20\% and 40\% of the machines in each node group for scheduling.
Restricting 40\% of the cluster shows that Tarema is able to decrease the runtime by 23.90\% compared to SJFN.
The log data analysis shows that SJFN sends the memory-intensive and long-running tasks, visualized in Figure~\ref{fig:workflow_profiles}, to the least powerful nodes.
Meanwhile, Tarema evenly distributes the tasks on the restricted resources.

\section{Conclusion}\label{sec:CONCLUSION}
This paper presented Tarema, a system that dynamically allocates scientific workflow tasks to heterogeneous cluster resources.
To this end, Tarema conducts a profiling run with microbenchmarks, creating detailed infrastructure performance profiles, and then clusters nodes with similar profiles into groups.
The system further labels the tasks of workflow jobs upon execution, thereby annotating resource demands without requiring separate task profiling runs.
Finally, Tarema uses the node groups and task labels to allocate tasks to available cluster resources at runtime.

We implemented a prototype of Tarema for the SWMS Nextflow, the resource manager Kubernetes, utilizing standard Linux benchmark tools.
Our evaluation with real-world scientific workflows reveals that through profiling, monitoring data, and task-resource profiles a geometric mean runtime decrease of 19.8\% over all workflows and runtimes can be achieved compared to standard schedulers for widely-used resource managers.
Since these schedulers are used in popular systems like Kubernetes, Yarn, or Slurm, this shows the improvements that can be achieved by adaptively allocating resources for workflows.
Further, through the profiling and monitoring extension, Tarema also enables the use of scheduling strategies like the heuristic SJFN.
Our comparison for isolated workflows with SJFN shows a geometric mean runtime decrease of 4.54\% over all workflows and runtimes, while the cluster usage using Tarema is better balanced over the different nodes.
Moreover, executing two workflows in parallel and on restricted resources shows that Tarema is able to reduce the runtimes even more while providing a fair cluster usage.

In the future, we plan to investigate how Nextflow can be extended to enable the support of multiple disk volumes to be able to allocate I/O intensive tasks more precisely.

\section*{Acknowledgments}
\thanks{Funded by the Deutsche Forschungsgemeinschaft (DFG, German Research Foundation) as FONDA (Project 414984028, SFB 1404).}

\bibliographystyle{IEEEtran}
\bibliography{./references}

\end{document}